\documentclass[reprint, superscriptaddress, amsmath,amssymb, aps,pra, longbibliography]{revtex4-1}

\usepackage{soul}
\usepackage{amsmath}   
\usepackage{amssymb}
\usepackage{graphicx}
\usepackage{dcolumn}
\usepackage{bm}
\usepackage{amsmath}
\usepackage{graphicx}
\usepackage{amsfonts}
\usepackage{subfigure}
\usepackage{graphicx}
\usepackage{array}
\usepackage{float}
\usepackage{color}

\usepackage{amssymb}%
\usepackage[colorlinks=true,linkcolor=blue]{hyperref}%
\hypersetup{
    allcolors=blue
}
\usepackage[normalem]{ulem}
\usepackage{xcolor}

\newcommand{\ket}[1]{\ensuremath{\left\vert #1 \right\rangle}}
\newcommand{\matrixel}[3]{\ensuremath{\left\langle #1 \middle| #2 \middle| #3 \right\rangle}}

\begin{document}

\title{Long-range Rydberg-atom-ion molecules of Rb and Cs}

\date{\today }

\author{A. Duspayev}
    \email{alisherd@umich.edu}
\affiliation{Department of Physics, University of Michigan, Ann Arbor, MI 48109, USA}
\author{X. Han}
\affiliation{Department of Physics, University of Michigan, Ann Arbor, MI 48109, USA}
\affiliation{State Key Laboratory of Quantum Optics and Quantum Optics Devices, Institute of Laser Spectroscopy, Shanxi University, Taiyuan 030006, People’s Republic of China}
\affiliation{Department of Physics, Taiyuan Normal University, Jinzhong 030619, People's Republic of China}
\author{M.A. Viray}
\affiliation{Department of Physics, University of Michigan, Ann Arbor, MI 48109, USA}
\author{L. Ma}
\affiliation{Department of Physics, University of Michigan, Ann Arbor, MI 48109, USA}
\author{J. Zhao}
\affiliation{State Key Laboratory of Quantum Optics and Quantum Optics Devices, Institute of Laser Spectroscopy, Shanxi University, Taiyuan 030006, People’s Republic of China}
\affiliation{Collaborative Innovation Center of Extreme Optics, Shanxi University, Taiyuan 030006, People's Republic of China}
\author{G. Raithel}
\affiliation{Department of Physics, University of Michigan, Ann Arbor, MI 48109, USA}

\begin{abstract}
We propose a novel type of Rydberg dimer, consisting of a Rydberg-state atom bound to a distant positive ion. The molecule is formed through long-range electric-multipole interaction between the Rydberg atom and the point-like ion. We present potential energy curves (PECs) that are asymptotically connected with Rydberg $nP$- or $nD$-states of rubidium or cesium. The PECs exhibit deep, long-range wells which support many vibrational states of Rydberg-atom-ion molecules (RAIMs). We consider photo-association of RAIMs in both the weak and the strong optical-coupling regimes between initial and Rydberg states of the neutral atom. Experimental considerations for the realization of RAIMs are discussed.
\end{abstract}

\maketitle
\section{Introduction} 

Ultralong-range Rydberg molecules (ULRM)~\cite{shafferreview, feyreview} have become an active field of research ever since they were first predicted~\cite{greene,boisseau} and experimentally observed~\cite{bendkowsky, overstreet}. There are two general types of ULRM. The first one is a bound state between ground-state and Rydberg-excited atoms. Its binding mechanism is a result of attractive interaction induced by the scattering of a Rydberg electron from an atom with negative scattering length, and is often described by the Fermi pseudopotential model~\cite{greene, fermi}. They exhibit a wide range of exotic properties such as macroscopic bond length, permanent electric dipole moment ranging from a few to thousands of Debye~\cite{li, booth, niederprum, tallant, bai2020} and unusual electronic probability densities~\cite{li,booth,niederprum, tallant, bellos}. Their photoassociation mechanisms~\cite{krupp, andersonprl, kleinbach}, scattering channels~\cite{sassmann, jamie} and interactions with external magnetic fields~\cite{lesanovsky, hummelpra2018, hummelpra2019} have been studied. Creation of Rydberg-ground molecules with alkaline-earth atoms~\cite{desalvo} and polyatomic Rydberg-ground molecules~\cite{bendkowskyprl, gaj, camargo} have been experimentally observed.
The second type of ULRM is formed by a pair of atoms that are both excited to Rydberg states, often referred to as Rydberg macrodimers~\cite{boisseau,  farooqi, overstreet}. In this case, the binding mechanism arises from the multipolar electrostatic interaction between the two atoms as has been studied in numerous theoretical works~\cite{samboy, samboyjpb, kiffner1, kiffner2, Han2019}. Rydberg macrodimers have been experimentally observed~\cite{overstreet, sassmannprl, Han2018} and recently studied using quantum gas microscopy~\cite{hollerith, hollerith2020}. \par

\begin{figure}[t]
 \centering
  \includegraphics[width=0.48\textwidth]{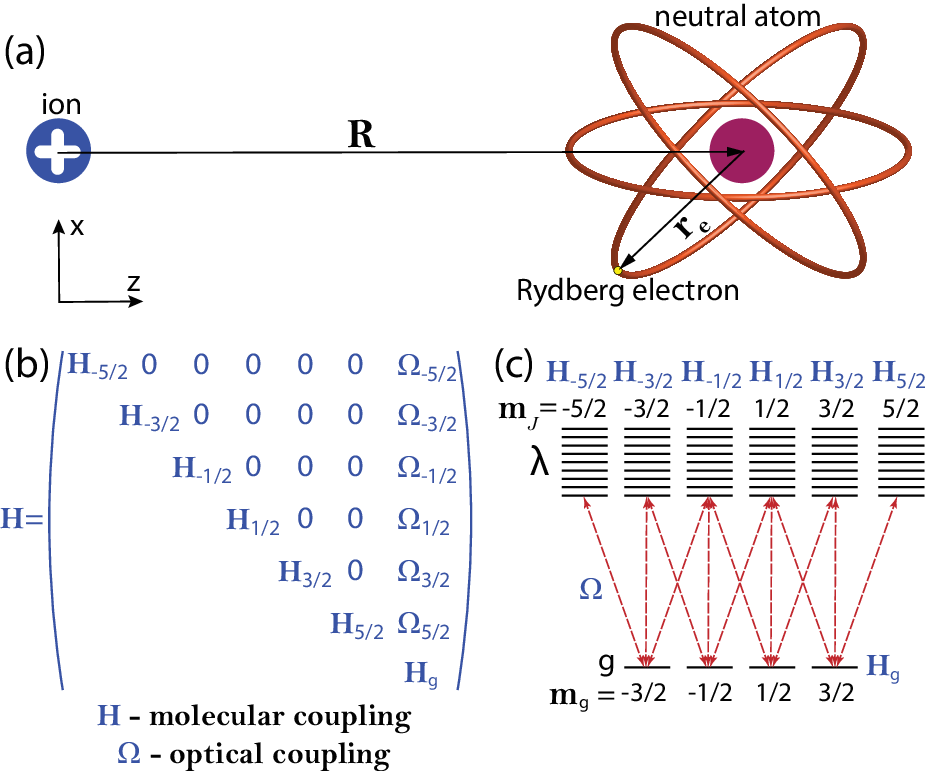}
  \caption{(Color online) (a) The Rydberg-atom-ion molecule (RAIM) under consideration. The molecular quantization axis and the internuclear distance $\mathbf{R}$ between the ion and the neutral atom are both along the \textit{z}-axis. The relative position of the Rydberg electron in the neutral atom is $\mathbf{r}_e$. (b) A block-matrix visualization of the Hamiltonian in Eq~\ref{eq:Hamiltonian}. (c) Energy-level schematic of the RAIM and optical couplings to a $J=3/2$ initial state, $\Omega$, with subspaces corresponding with the notations as defined in~(b). }
  \label{figure 1}
\end{figure}

While Rydberg-ground and Rydberg-Rydberg molecules have been of great interest, experimental implementation of hybrid atom-ion systems~\cite{harter, tomza} also have led into opportunities to realize a different type of ULRM - one formed due to an interaction between a cold Rydberg atom and a trapped ion. Hybrid atom-ion systems are generally attractive for novel quantum chemistry at ultracold temperatures~\cite{ospelkaus, schlagmuller, TSchmid2018}, many-body physics~\cite{cote2002, casteels, joger, dieterle, hirzler, dieterleprl, benschlomi}, quantum sensing and control~\cite{engel, demille, carr, ewald} as well as quantum computing and simulations~\cite{doerk, seckerpra, bissbort}. However, atom-ion collisions can lead to depletion of the ultracold atomic ensemble, micromotion-induced collisions, charge transfer and other unwanted effects~\cite{zipkesnature, zipkesprl, schmid}. Among methods to prevent these processes, a tight confinement of atoms and ions in individual traps at sufficiently large distances~\cite{idziaszek} and optical shielding~\cite{secker, wang} have been suggested. The latter method is based on exciting neutral atoms colliding with ions into certain Rydberg states that become repelled. The method also lends itself to the preparation of Rydberg-atom-ion molecules (RAIMs). \par 

Here we investigate RAIMs between Rydberg states of rubidium and cesium atoms and point-like positive ions. Similar to the aforementioned Rydberg macrodimers, the adiabatic potential energy curves (PECs) of RAIMs arise from non-perturbative solutions of the atom-ion interaction Hamiltonian. Bound vibrational states exist within sufficiently deep wells in PECs that are not coupled to un-bound PECs. After outlining the theoretical model in Sec.~\ref{sec:model}, in Sec.~\ref{sec:results} we calculate the PECs of the atom-ion systems and identify series of bound vibrational states of RAIMs near the intercepts between PECs that asymptotically connect with $nP_J$ Rydberg states and with hydrogenic states (states with near-zero quantum defect). We assess the convergence of the PECs as a function of the highest Rydberg-atom multipole order included in the calculation. We study RAIM systems in cases of both weak and strong optical coupling between initial and Rydberg states in the neutral atom. The initial state can be the atomic ground state or another low-lying intermediate atomic state. Implications regarding spectroscopic measurement of RAIMs and coherent vibrational dynamics are discussed, as well as the context with optical-shielding applications~\cite{secker, wang}. While RAIMs exist over a wider range of $n$, here we show results for Rb and Cs RAIMs with Rydberg levels between $n=45$ and~55, because these levels are easily laser-excitable and their ionization electric fields~\cite{gall},   $\sim$50~V/cm, are low enough for convenient RAIM detection. Also, their RAIM bonding lengths and vibrational frequencies are conducive to experimental study. In Sec.~\ref{sec:expt} we discuss RAIMs from several experimental points of view. The paper is concluded in Sec.~\ref{sec:conclusion}.

\section{Theoretical model} 
\label{sec:model}

\subsection{Hamiltonian and potential energy curves}
\label{subsec:hamiltonian}
The system under consideration is shown in Fig.~\ref{figure 1}(a). We consider a Rb or Cs Rydberg atom near a positive ion at an internuclear separation $\bf{R}$. The molecular quantization axis $\hat{\bf{z}}$ is chosen to point along $\bf{R}$. The relative position of the valence electron is $\mathbf{r}_e=(r_e, \theta_e, \phi_e)$. The internuclear distance $R$ is larger than the radius of the Rydberg atom, such that dissociation through a process Rb$^*_{n>}$Rb$^+ \longrightarrow $Rb$^*_{n<} + $Rb$^+$ is avoided. This is similar to the Le Roy radius condition~\cite{LeRoy} in Rydberg macrodimers~\cite{boisseau}. Also, there is no radiation retardation effect~\cite{Casimir}, because the energy differences in the Rydberg-atom Hilbert space correspond with transition wavelengths much larger than $R$.  Further, the ion is considered as a positive point charge without relevant internal substructure. This approximation is applicable for ions such as a Rb$^+$ or Cs$^+$, in their electronic ground state. In these cases, level shifts of the atom-ion pair due to dipolar ionic polarization at distances $\gtrsim 1~\mu$m, considered here, range in the sub-Hz regime, rendering ion polarization effects negligible. Since the ion has no internal substructure, the Hilbert space is that of one Rydberg atom, as opposed to a product space of two such atoms (as in Rydberg-Rydberg molecules~\cite{boisseau, samboy, samboyjpb, kiffner1, kiffner2, farooqi, overstreet, sassmannprl, Schwettmann2006, hollerith, Han2018, Han2019, hollerith2020}).\par

The Rydberg-state Hilbert-space used is $\{\ket{\lambda, m_{J}}\}$, where the index $\lambda$ is a shorthand for the $n$, $\ell$ and $J$ quantum numbers of the Rydberg base kets. The magnetic quantum number $m_J$ is defined in the molecular reference frame. In the absence of strong optical coupling, $m_J$ is conserved due to azimuthal symmetry. The total Hamiltonian of the neutral atom - ion system then is, in the molecular frame,
\begin{eqnarray}
\begin{aligned}
\hat{H} (R) = \hat{H}_g + \sum_{m_J} \left( \hat{H}_{m_J}(R) + \left\{ \hat{\Omega}_{m_J,g} + c.c. \right\}\right).\qquad
\end{aligned}
\label{eq:Hamiltonian}
\end{eqnarray}
\noindent Here, $\hat{H}_g$ is the Hamiltonian of the neutral atom in the initial-state subspace $\{\ket{\lambda_g, m_g} \}$. There, $\lambda_g$ denotes the fixed quantum numbers $n_g$, $\ell_g$ and $J_g$ of the initial state, which is treated as an effective ground state, and $m_g$ its magnetic quantum number in the lab (laser) frame. The initial-state subspace includes all $m_g = -J_g, ..., J_g$. The sum in Eq.~\ref{eq:Hamiltonian} is taken over Rydberg-atom Hamiltonians $\hat{H}_{m_J}$, which act on Rydberg-atom subspaces $\{\ket{\lambda, m_{J}}\}$. Each Rydberg-atom subspace has a fixed value of $m_J$, in the molecular frame, and a range of $\lambda$'s covering all Rydberg states with that $m_J$ and within a specified energy interval. The optical couplings between the initial-state levels and Rydberg states are introduced via the operators $\hat{\Omega}_{m_J, g}$, the matrix elements of which are optical excitation Rabi frequencies between Rydberg subspace $\{\ket{\lambda, m_{J}}\}$ and initial-state subspace $\{\ket{\lambda_g, m_g} \}$. The Rabi frequencies depend on magnitude and polarization of the field that couples initial-state and Rydberg manifolds, and on the geometry of the excitation laser relative to the molecular reference frame.  \par

In our calculations we use initial states $5D_{3/2}$, $5P_{3/2}$ or $5S_{1/2}$ for Rb, and $6D_{3/2}$ or $6S_{1/2}$ for Cs, which all allow optical excitation of field-mixed Rydberg $nP_J$ states. The hyperfine structure of the initial state is neglected, because it does not add additional insight to the physics presented in this paper. Also, $\hat{H}_g$ does not depend on $R$, because the effect of the ion electric field on our initial states is negligible. The optically coupled Rydberg-atom subspaces $\{\ket{\lambda, m_{J}}\}$ included in Eq.~\ref{eq:Hamiltonian} cover the range $m_J = -J_g-1, ..., J_g+1$. The corresponding $\hat{H}_{m_J}$ can be written as:
\begin{eqnarray}
\hat{H}_{m_J}(R) = \hat{H}_{Ry, m_J} + V_{int, m_J}(\mathbf{\hat{r}}_e; R),
\label{eq:H_i}
\end{eqnarray}
\noindent where $\hat{H}_{Ry, m_J}$ is the unperturbed Hamiltonian of the manifold of Rydberg states with magnetic quantum number $m_J$ in the molecular frame. For the quantum defects of the atomic energies and the fine structure coupling constants in $\hat{H}_{Ry, m_J}$ we use previously published values~\cite{gall,Seaton}. The operator $\hat{V}_{int, m_J}$ denotes the multipole interaction between the Rydberg atom and the ion, which is $m_J$-conserving and is described below. \par 

In Fig.~\ref{figure 1}(b), we show a block-matrix representation of the Hamiltonian in Eq.~\ref{eq:Hamiltonian}. The structure of the Hilbert space is visualized with the energy level diagram in Fig.~\ref{figure 1}(c). The block-diagonal structure evident in Fig.~\ref{figure 1}(b) is due to the azimuthal symmetry of the multipolar interaction between the Rydberg electron and the ion within the molecular reference frame. \par

The PECs without optical coupling or with weak optical coupling follow from diagonalizations of Eq.~\ref{eq:H_i} within the subspaces $\{\ket{\lambda, m_{J}}\}$ for fixed $m_J$. The procedure is performed on a dense grid of the internuclear separation $R$, and for all desired choices of $m_J$. The interaction $\hat{V}_{int, m_J}$ is~\cite{Schwettmann2006,Deiglmayr2014,Han2018} (in atomic units):
\begin{eqnarray}
V_{int, m_J} (\mathbf{\hat{r}}_e; R) = - \sum_{l=1}^{l_{max}}\sqrt{\frac{4\pi}{2l+1}}\frac{\hat{r}_e^{l}}{R^{l+1}}Y_{l 0}(\hat{\theta}_e, \hat{\phi}_e)
\label{eq:Vin}
\end{eqnarray}
\noindent Here $l$ is the multipole order of the Rydberg atom, the position operator $\mathbf{\hat{r}}_e$ is the relative position of the Rydberg electron in the neutral atom, and $Y_{l 0}(\hat{\theta}_e, \hat{\phi}_e)$ are spherical harmonics that depend on the angular position of the Rydberg electron. The basis is truncated in energy by setting lower and upper limits of the effective quantum number, $n_{\rm{min}} \leq n_{\rm{eff}} \leq n_{\rm{max}}$, typically covering a range of $n_{\rm{max}} - n_{\rm{min}}$ between 5 and 10. The interaction in Eq.~\ref{eq:Vin} conserves $m_J$, with representations in different subspaces $\{\ket{\lambda, m_{J}} \, \big| \, m_J = {\rm{const}}\}$ being different (which is why we use a subscript $m_J$ on $\hat{V}_{int,m_J}$). The adiabatic molecular states of the RAIMs without optical coupling are $\ket{k, m_J} = \sum_{\lambda} c_{\lambda, m_J, k} \ket{\lambda, m_J}$, with conserved $m_J$, and the index $k$ serving as a counter for PECs of the RAIM states. \par 
 
The sum over the multipole order $l$ in Eq.~\ref{eq:Vin} starts at 1, because the Rydberg atom has no electric monopole moment, and is truncated at a maximal order $l_{max}$. The selection rules for the matrix elements $\langle n', \ell', J', m'_J \vert \hat{V}_{int, m_J} \vert n, \ell, J, m_J \rangle $ are $\vert \ell' - \ell \vert = l_{max}, l_{max} -2, ..., 0$~or~1, and $\vert J' - J \vert \leq l_{max}$ and $m'_J = m_J$. Since all atomic states with $\ell < n$ are included in the Hilbert space anyways, a larger $l_{max}$ only increases the number of non-zero interaction matrix elements, but does not increase the dimension of the state space. Here we use $l_{max}=6$, which is sufficient for convergence at a practical level. \par 

Since the quantization axis is given by the internuclear separation vector, mixing of different $m_J$ states only occurs via coherent, laser-induced electric-dipole coupling through the initial-state subspace $\{\ket{\lambda_g, m_g}\}$. The optical coupling matrix elements are contained in the off-diagonal $\Omega_{m_J}$-blocks in Fig.~\ref{figure 1}(b) and are visualized by arrows in Fig.~\ref{figure 1}(c). In the following Sub-sections, we consider the distinct cases of weak and strong optical coupling. \par

In the regime of weak optical coupling, discussed in Secs.~\ref{subsec:weakfield} (theory) and~\ref{subsec:weakfieldres} (results), the Rabi frequencies $\Omega$ between initial-state levels and Rydberg states are negligible against Rydberg- and initial-state decay (and possibly other sources of decoherence). In this case, coherences and optically-induced splittings between PECs and RAIM states with different $m_J$ are negligible. This regime corresponds to conditions suitable for spectroscopic measurements aimed at experimental observation of RAIMs and resolving their vibrational states.

In the regime of strong optical coupling, discussed in Secs.~\ref{subsec:strongfield} (theory) and~\ref{subsec:strongfieldres} (results), laser intensities are high enough that the optically induced couplings between manifolds of Rydberg states with different $m_J$ are not negligible, or may even become dominant at certain internuclear separations $R$ and molecular energies. One then finds modifications of the PECs and changes in the molecular dynamics. The strong-optical-coupling regime covers applications such as optical shielding~\cite{secker, wang}. Also, the study of coherent vibrational wavepackets can require laser excitation pulses that are in that regime. \par

It is, generally, noted that the multipolar interaction of a Rydberg atom with an ion is less complex than that between a Rydberg-atom pair~\cite{Schwettmann2006,Deiglmayr2014,Han2018}. Since ionic multipoles beyond the monopole are negligible, the calculation resembles that of a Stark-effect calculation for a Rydberg atom in an electric field, with the addition of a large number of non-dipolar matrix elements that account for the interaction of the atom with the inhomogeneity of the ion electric field. The Hilbert space remains, however, a relatively small single-atom space, as opposed to a product space of two atoms~\cite{Schwettmann2006,Deiglmayr2014,Han2018}. As a consequence, in the theory of RAIMs there are no significant concerns about errors caused by basis-size truncation.

\subsection{Weak optical coupling }
\label{subsec:weakfield}

If the elements of the six $\Omega_{m_J}$-blocks in Fig.~\ref{figure 1}(b), which represent the optical Rabi frequencies from the initial state, are small relative to the decay- and decoherence rates of the system, we separately diagonalize the six $H_{m_J}$-blocks in Fig.~\ref{figure 1}(b) to obtain the PECs and adiabatic molecular states for the desired $m_J$-value(s). The calculation yields the PEC energies, $W_{k, m_J}(R)$, of the molecular states $\ket{k, m_J}$, and their excitation rates, $T_{k, m_J}$. The latter are calculated using Fermi's golden rule:    
\begin{eqnarray}
T_{k, m_J}(R) = \frac{2\pi}{\hbar} \sum_{m_g=-J_g}^{J_g} 
\rho \, p(m_g)\,\, S(k, m_J, m_g) \quad,
\label{eq:excrates}
\end{eqnarray}
\noindent which depend on the PEC indices $k$ and $m_J$, the internuclear separation $R$, and other quantities as follows. We assume that the laser bandwidth is the dominant broadening source. In this case, the energy density of states, $\rho$, is approximately given by the inverse of the FWHM energy bandwidth of the excitation laser (for all PECs).  The $p(m_g)$ are statistical weights for the atom populations in the magnetic sublevels of the initial state, $m_g$, which are given in the lab (excitation-laser) frame of reference. The signal strengths $S(k, m_J, m_g)$ are related to the optical transition matrix elements
\begin{eqnarray}
A(k, m_J, m_g, \theta) = \matrixel{k, m_J}{\hat{D}_y(\theta)~\hat{\Omega}}
{\lambda_g, m_g} 
\quad .
\label{eq:excelements}
\end{eqnarray}
\noindent Here, $\hat{D}_y(\theta)$ is the operator used for rotating from the laboratory frame, defined by the excitation-laser geometry, into the molecular frame, defined by the internuclear axis. Eq.~\ref{eq:excelements} can be explicitly written as:
\begin{eqnarray}
A =  \sum_{\lambda} \sum_{\tilde{m}} \, c_{\lambda, m_{J},k}^* \, 
d^{(J)}_{m_{J}, \tilde{m}}(\theta) \, \langle \lambda, \tilde{m} \vert
E_L \hat{\bf{\epsilon}} \cdot \hat{\bf{r}}_e \vert \lambda_g, m_g \rangle 
\label{eq:excelementsexplicit}
\end{eqnarray}
\noindent with the excitation-laser electric field, $E_L$, and its polarization vector in the lab frame, $\hat{\bf{\epsilon}}$. The first sum is going over the base Rydberg states $\ket{\lambda, m_J}$, with the shorthand $\lambda$ for the Rydberg-state quantum numbers $n$, $\ell$ and $J$, and the second over a Rydberg-state magnetic quantum number $\tilde{m}$ in the lab frame. The electric-dipole matrix element is between a Rydberg state  $\vert \lambda, \tilde{m} \rangle$ and an initial-state, $\vert \lambda_g, m_g \rangle$,  with both magnetic quantum numbers $\tilde{m}$ and $m_g$ in the lab frame. The rotation from the lab into the molecular frame is affected by elements of reduced Wigner rotation matrices, $d^{(J)}_{m_J, \tilde{m}}$. \par

The optical transition matrix elements $A(k, m_J, m_g, \theta)$ given by Eq.~\ref{eq:excelementsexplicit} depend on $\theta$ through the rotation operator matrix elements $d^{(J)}_{m_J,\tilde{m}}(\theta)$, which rotate the optically excited Rydberg level $\vert \lambda, \tilde{m} \rangle$ from the lab into  the molecular frame. The excitation 
signal strength $S(k, m_J, m_g)$ of the PEC for $\ket{k, m_J}$, from the initial-state level  $\vert \lambda_g, m_g \rangle$  then is a weighted average of Eq.~\ref{eq:excelementsexplicit} over the alignment angle $\theta$ of the molecules relative to the lab frame:
\begin{equation}
    S(k, m_{J}, m_g) \propto\int_0^\pi P(\theta)|A(k, m_{J}, m_g, \theta)|^2\ \mathrm{d}\theta
 \label{eq:theta_avg}
\end{equation}
where $P(\theta)$ is the probability weighting function. For an isotropic atomic sample with random molecular alignment:
\begin{equation}
    P(\theta)=\frac{\sin(\theta)}{2} \quad.
\label{eq:theta_weight_func}
\end{equation}
In the presented excitation signal-strength calculations for the case of weak optical coupling, we assume a sample with random molecular alignment, and with equal populations of $p(m_g) = 1/(2 J_g +1)$ in all magnetic sublevels of the initial state. Equations~(\ref{eq:excrates}) to~(\ref{eq:theta_weight_func}) yield the excitation rates $T_{k, m_J}(R)$ of the PECs for $\ket{k, m_J}$.

\subsection{Strong optical coupling}
\label{subsec:strongfield}

In the case of strong optical coupling, the Rabi frequencies contained in the $\Omega_{m_J}$-blocks in Fig.~\ref{figure 1}(b) become larger than decay and decoherence of initial and Rydberg levels. Strong  coupling is, for instance, required for optical shielding~\cite{secker, wang}, where the passage behavior of the Rydberg atom - ion system through optically induced anti-crossings has to be adiabatic for the shielding to be effective. In the cases studied in the present paper, the optical Rabi frequencies necessary for strong coupling are $\gtrsim 10$~MHz. \par 

In the strong-coupling case, the matrix elements in the $\Omega_{m_J}$-blocks are given by the electric dipole couplings
\begin{eqnarray}
~ & \langle \lambda, m_J \vert \hat{\Omega}_{m_J, g} \vert \lambda_g, m_g \rangle 
 = & ~  \nonumber \\
~ & \sum_{\tilde{m}} d^{(\lambda)}_{m_J, \tilde{m}} (\theta)  \langle \lambda, \tilde{m} \vert
E_L \hat{\bf{\epsilon}} \cdot \hat{\bf{r}}_e \vert \lambda_g, m_g \rangle & ~  
\label{eq:strong}
\end{eqnarray}
with $m_J$ in the molecule frame, and $m_g$ and $\tilde{m}$ in the lab frame. The optical coupling has up to three components ($\sigma^{+/-}$ and $\pi$), corresponding to different values of $\tilde{m}-m_g$. The matrix elements are now explicitly included in the $\Omega_{m_J}$-blocks of the total Hamiltonian in Fig.~\ref{figure 1}(b). Due to the rotation, generally all matrix elements in the $\Omega_{m_J}$-blocks are different from zero. In effect, all Rydberg states that correspond with the different $H_{m_J}$-blocks become coupled to each other in second order of the optical interaction. Hence, in the regime of strong optical coupling the entire Hamiltonian in Fig.~\ref{figure 1}(b) has to be diagonalized.\par 

In our calculation, the Hamiltonian is expressed in an interaction picture, with a number of $2 J_g+1$ dressed~\cite{note1} initial-state levels, $\vert \lambda_g, m_g \rangle$, intersecting with Rydberg PECs, $\vert k, m_J \rangle$, at energies and internuclear separations that depend on the excitation-laser  detuning from the Rydberg $nP_J$-levels. The diagonalization of the full Hamiltonian of Fig.~\ref{figure 1}(b) yields optically perturbed and mutually coupled PECs of all $m_J$'s at once. Generally, the PECs belonging to different $m_J$'s become coherently mixed in second order of the optical coupling via the initial-state sublevels. The mixing tends to be fairly localized in $R$ and energy (see Fig.~\ref{figure 4} below), because the optical couplings are typically much smaller than the energy variation of the PECs.  

\begin{figure*}[t]
 \centering
  \includegraphics[width=\textwidth]{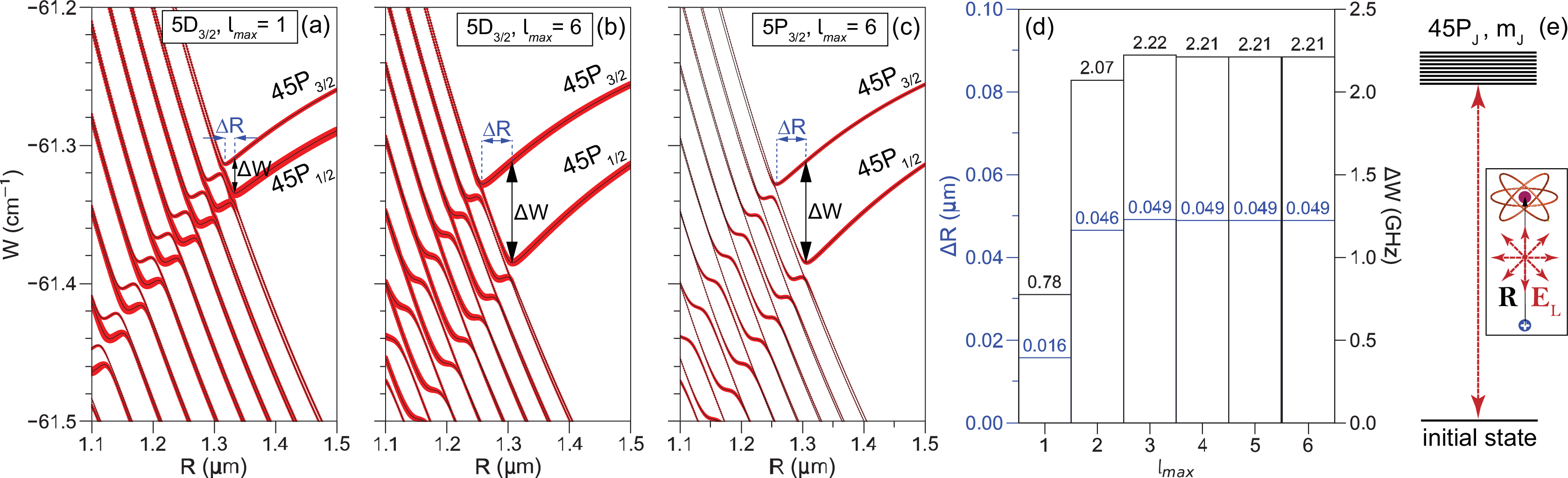}
  \caption{Rydberg-atom-ion molecules (RAIMs) for a case in rubidium. (a) PECs as a function of internuclear separation, $R$, calculated for $l_{max} = 1$ in Eq.~\ref{eq:Vin}. The plotted PECs are relative to the atomic ionization potential.
  The width of the red backdrop lines is proportional to the single-atom PA rate for the excitation geometry shown in panel (e) and details explained in the text. The PA rate on the outer slope of the lower PEC well is $\approx 3.7 \times 10^5$~s$^{-1}$. (b) Same as panel (a), but with $l_{max} = 6$. Deviations from panel (a) are due to higher-order multipolar effects associated with the inhomogeneity of the ion's electric field.  The PA rate on the outer slope of the lower PEC well is $\approx 2.6 \times 10^5$~s$^{-1}$. (c) Same as panel (b), but with the PA rates calculated for excitation from the $5P_{3/2}$ initial state. The PA rate on the outer slope of the lower PEC well is $\approx 0.61 \times 10^5$~s$^{-1}$. (d) Effect of $l_{max}$ in Eq.~\ref{eq:Vin} on the PEC fine-structure splitting, $\Delta W$, and the differential internuclear separation, $\Delta R$, as defined in panels (b) and (c), between the two most prominent minima in the relevant PECs ($\Delta W$ and $\Delta R$ do not depend on initial state). (e) Qualitative energy-level diagram and excitation-laser geometry used for the PA-rate calculations in panels (a)-(c).}
  \label{figure 2}
\end{figure*}

\section{Results and discussion}
\label{sec:results}

\subsection{Determination of $l_{max}$}
\label{subsec:lmax}

In Figs.~\ref{figure 2}~(a) and~(b), we show RAIM PECs calculated within the Rydberg subspace $\{\ket{\lambda, m_{J}} \, \big| \; 38.9 \leq n_{\rm{eff}} \leq 46.1 
\; {\rm{and}} \; \ell = 0, 1, ..., n-1
\; {\rm{and}} \; J = \ell \pm 1
\; {\rm{and}} \; m_J = 1/2
\} $ of Rb as a function of the internuclear separation, $R$, for $l_{max}=1$ and 6 in Eq.~\ref{eq:Vin}, respectively. This subspace is centered around the atomic states $45P_J$. For $l_{max}=1$, the plot is identical with the diagram of Stark states of Rb in a homogeneous electric field, $E$, given by the ion's electric field at the Rydberg-atom center, $E=1/R^2$ (in atomic units). For both values of $l_{max}$, the PECs exhibit well-defined minima in the intersection regions between the ion-electric-field-mixed $45P_J$-PECs coming in from $R=1.5~\mu$m (marked in Fig.~\ref{figure 2}), and a fan of many slightly curved, approximately parallel PECs in the lower-left halves of the plots. For $l_{max}=1$, the latter are identical with the set of linear Stark states~\cite{gall} of the $n=42$ manifold of Rb. The absence of narrow anti-crossings in the lower potential well (at the lower tips of the arrows) indicates the existence of bound RAIM states that are stable against decay via non-adiabatic coupling to unbound PECs. The observed potential minima are deep enough to support several tens of vibrational states (see Sec.~\ref{subsec:strongfieldres}). \par

Comparing Figs.~\ref{figure 2}~(a) and~(b), we observe a significant influence of $l_{max}$ on the calculated PECs. This is due to the strong inhomogeneity of the ion electric field, which necessitates the inclusion of several higher-order multipole contributions in Eq.~\ref{eq:Vin}. Several qualitative differences between the results for $l_{max}=1$ in Fig.~\ref{figure 2}(a) and $l_{max}=6$ in Fig.~\ref{figure 2}(b)) are apparent. Including more terms in Eq.~\ref{eq:Vin} leads to an increase in $\Delta W$, defined as the splitting between the PECs that asymptotically connect with the $nP_J$ fine-structure components at the location of the relevant PEC potential minima. The minima of the PEC wells that asymptotically connect with the $45P_{1/2}$ and $45P_{3/2}$ atomic states occur at different internuclear distances, separated by a difference $\Delta R$, as seen in Fig.~\ref{figure 2}(b). In Fig~\ref{figure 2}(d) we show the convergence of $\Delta W$ and $\Delta R$ as a function of the number of multipole orders included. One can see that the inclusion of the $l=2$ term is particularly important; this is because the coupling between the $45P_J$ fine-structure components is $l=1$-forbidden but $l=2$-allowed. Indeed, the energy scale of the matrix elements in Eq.~\ref{eq:Vin} between $nP_{1/2}$ and $nP_{3/2}$ states near the relevant PEC potential minima is on the same order of magnitude as the fine structure splitting between these states, and therefore generates significant level repulsion. Thus, the quadrupolar term, $l=2$, is particularly important to include, because it causes level repulsion between the $nP_J$ fine-structure components. This has a large effect on both $\Delta W$ and $\Delta R$. Successively higher orders have increasingly smaller effects, as seen in Fig~\ref{figure 2}(d). In our further calculations we set $l_{max} = 6$, allowing for sub-MHz and sub-nm accuracy in the energies and locations of the minima on the PEC surfaces, respectively.

\subsection{Weak optical coupling}
\label{subsec:weakfieldres}

After confirming that the emergence of the RAIMs is robust against $l_{max}$ in Eq.~\ref{eq:Vin}, we discuss how one could probe RAIMs optically. We first consider the excitation of Rb RAIMs in the regime of weak optical coupling, as described in Sec.~\ref{sec:model}. The atoms are initially in the initial state $5D_{3/2}$ (Figs.~\ref{figure 2}~(a) and~(b)) or $5P_{3/2}$ (Fig.~\ref{figure 2}(c)), and are located at a distance $R$ from a Rb$^+$ ion. The photo-association (PA) laser is assumed to be linearly polarized and to have an intensity of $10^9$~W/m$^2$, corresponding to a field amplitude of $868 \times 10^3$~V/m. The angle $\theta$ between $\textbf{E}_L$ and the molecular axis is random, as visualized in the inset in Fig.~\ref{figure 2}(e), and the initial atoms are evenly distributed over the magnetic sub-states $\vert \lambda_g, m_g \rangle$. The width of the red backdrop lines behind the PECs in Figs.~\ref{figure 2}(a)-(b) is proportional to the average PA rate for a single atom, calculated by means of Eqs.~\ref{eq:excrates}-\ref{eq:theta_weight_func}. The energy density of states for resonant PA is $\rho = 1 / ( \sqrt{2 \pi} h \sigma_L)$, with a full width at half maximum of the laser spectral intensity vs frequency of $\sigma_L \sqrt{8 \ln 2} = 1$~MHz. \par

Most importantly, it is seen that the PA rates are high on the PECs that form the potential wells, indicating feasibility of exciting and observing RAIMs via excitation from low-lying atomic levels. The PA rates are high on the outer slopes of the PEC wells, where the adiabatic states have substantial $P$-character, while the rates are small on the inner slopes of the wells, where the adiabatic states resemble linear Stark states with small low-$\ell$ character. The electric-dipole selection rules for the laser excitation then largely explain the distribution of oscillator strength over the PECs. The essential physics here is similar to that of the Stark effect in alkali Rydberg atoms with large quantum defects (Rb and Cs)~\cite{Zimmerman.1979}. Since the PA rates in the PEC wells are strongly lopsided towards the outer slopes, almost all PA of vibrational RAIM states will occur near the outer turning points of the wells. This fact is expected to enable initialization of vibrational RAIM wavepackets at the outer turning point via pulsed PA~(see Sec.~\ref{subsec:wavepacketexp}). \par  

Further examination of Figs.~\ref{figure 2}~(a) and~(b) shows that for $l_{max}=6$ the PA rates into the PECs asymptotically connected with the $45P_J$ states become redistributed between the $J=1/2$ and 3/2 states, and become near-identical. The redistribution does not occur in the case $l_{max}=1$. This qualitative difference in behavior results from the fact that the $l=2$-terms in Eq.~\ref{eq:Vin} strongly mix the fine-structure components and thereby tend to average out their oscillator strengths. If the $l=2$-terms were left out, such as in Fig.~\ref{figure 2}(a), the calculation fails to account for the redistribution of oscillator strength between the $nP_J$ fine-structure pairs. \par

Since the ion electric field mixes some $S$ and $D$-character into the PECs that asymptotically connect with the $nP_J$ states, PA of RAIMs from $5P_{3/2}$ as an initial state also is fairly efficient. This is shown in Fig.~\ref{figure 2}(c), where we employed Eqs.~\ref{eq:excrates}-\ref{eq:theta_weight_func} to compute the PA rates from $5P_{3/2}$. For the PA laser intensity and other conditions identical with those in Fig.~\ref{figure 2}(b), for the $5P_{3/2}$ initial state we expect a PA rate that is about 1/5 of that for the $5D_{3/2}$ initial state. Comparing Figs.~\ref{figure 2}~(b) and~(c), it is further noted that the lopsidedness of the PA rates between the inner and outer slopes in the PEC wells is greater for the $5P_{3/2}$ initial state than it is for $5D_{3/2}$. Finally, since the decay rate of $5P_{3/2}$ is about ten times higher than that of $5D_{3/2}$, when using $5P_{3/2}$ as an initial state one would consider two-photon, off-resonant PA of RAIMs in order to reduce or eliminate decay and decoherence from the $5P_{3/2}$-decay.

\subsection{Comparison of Rb and Cs}
\label{subsec:rbvscs}

\begin{figure}[t!]
 \centering
  \includegraphics[width=0.49\textwidth]{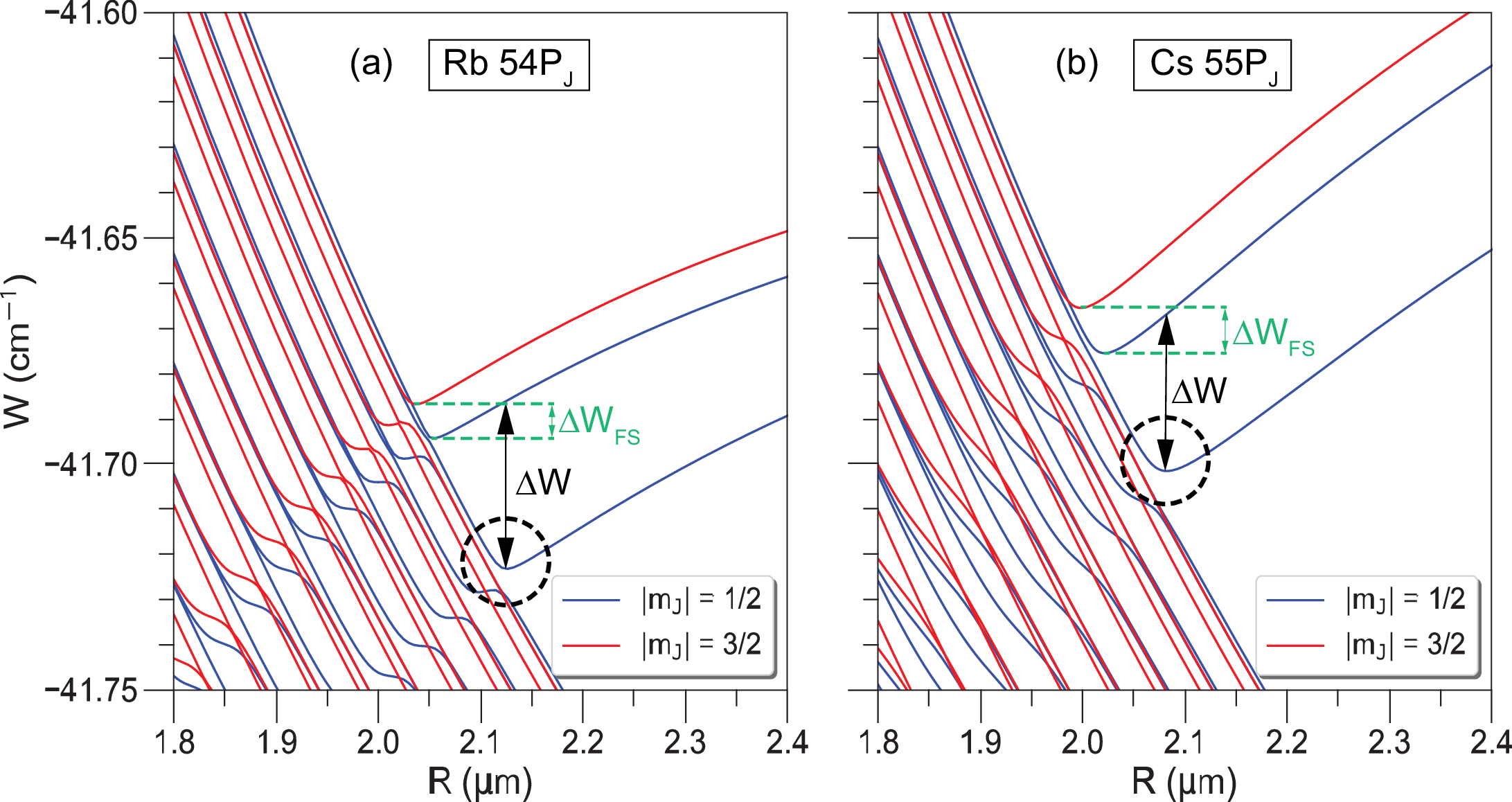}
  \caption{PECs relative to the atomic ionization potentials as a function of internuclear distance $R$, for rubidium (a) and cesium (b). Both atomic species exhibit similar RAIMs, with mild variations in potential depth, fine-structure splitting and bond length.}
  \label{figure 3}
\end{figure}

Before we proceed to discuss the strong optical coupling regime, in this section we provide a comparison between Rb and Cs. We calculate the PECs of RAIMs near the $54P_J$ and $55P_J$ Rydberg states of Rb and Cs (Figs.~\ref{figure 3}(a) and Fig.~\ref{figure 3}(b), respectively). These states allow for a comparison at similar energies, because the $P$ quantum defects in Rb and Cs differ by $\sim 1$. The general shapes of the PECs neither depend on $n$ nor on type of atom (Cs or Rb). A key finding therefore is that the RAIM bound states are expected to be observable near $nP_J$ Rydberg states in both Rb or Cs, over a wide range of $n$. Due to the similarities in the non-integer parts of the quantum defects, both cases exhibit PEC minima at similar locations within the spectra. Hence, similar RAIMs are predicted to exist in both Rb and Cs. \par

We find several minor differences, which result from the different quantum defects and ﬁne-structure coupling constants of Rb and Cs. First, the depth $\Delta W$ of the lower wells, marked by the dashed circles in Fig.~\ref{figure 3},  relative to the next-higher PECs is slightly deeper in Rb than in Cs (by $\sim 8\%$). Second, the binding length of the circled potential well for Cs ($2.08\mu$m) is slightly smaller than that of Rb ($2.12\mu$m). Further, the splittings $\Delta W_{FS}$ between the close-by pairs of PEC wells for $\vert m_{J} \vert =1/2$ and $\vert m_{J} \vert =3/2$ are larger in Cs than they are in Rb (see $\Delta W_{FS}$-marks in Fig.~\ref{figure 3}), which is due to the larger fine-structure coupling in Cs. Finally, the curvature of the circled PEC well in Cs is less than that in Rb, indicating stronger coupling and level repulsion between the $P$-like and the linear-Stark-state-like PECs in Cs than in Rb. This means that non-adiabatic decay of bound vibrational states into dissociating states on the PECs that resemble linear Stark states is less likely in Cs than it may be in Rb.

Generally, the existence of RAIM states of the type discussed here is related to the existence of low-$\ell$ Rydberg states that red-shift in an external electric field, equivalent to a positive DC polarizability, and that intersect with blue-shifting linear Stark states~\cite{Zimmerman.1979, gall}. Further, sufficient coupling between the low-$\ell$ and linear Stark states is required, so as to suppress non-adiabatic coupling of bound RAIM states to states on dissociating PECs. The existence of these features relies on the low-$\ell$ quantum defects, which have to be sufficiently large to produce substantial coupling between low-$\ell$ and linear Stark states, and which must have non-integer parts between $\sim 0.5$ and $\sim 0.8$. Suitable cases include alkali atoms of K, Rb and Cs, but there likely exist other cases among a wider variety of Rydberg atoms.

\subsection{Strong optical coupling}
\label{subsec:strongfieldres}
\begin{figure}[t!]
 \centering
  \includegraphics[width=0.42\textwidth]{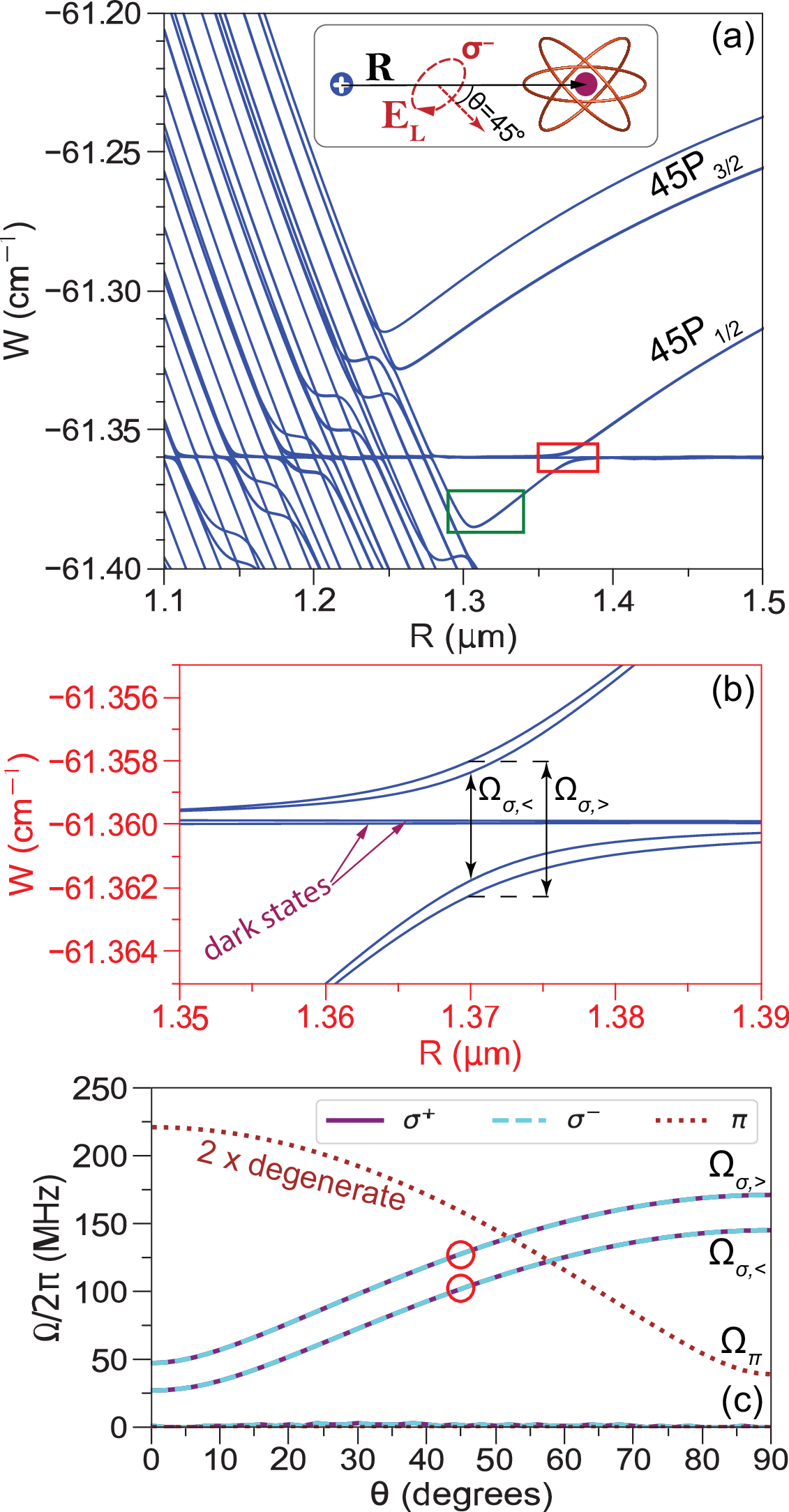}
    \caption{Rydberg-atom-ion molecules (RAIMs) of rubidium in the case of the strong optical coupling. (a) Dressed initial-state and RAIM PECs as a function of internuclear distance $R$, for a fixed atom-field detuning of 5.957~GHz relative to the field-free atomic $45P_{3/2}$ resonance. The system is driven by $\sigma^-$-polarized light propagating at $\theta=45^\circ$ with respect to the molecular quantization axis as shown in the inset and with intensity of $\approx 1.33\times10^9$~W/m$^2$ corresponding to electric-field amplitude of $E_L=1\times10^6$~V/m. (b) Magnified view of the region of intersection between the dressed PECs of the initial-state and the RAIM PEC that asymptotically connects with the $45P_{1/2}$ state. Optical Rabi frequencies are indicated by vertical arrows. (c) Optical Rabi frequencies for linear ($\Omega_{\pi}$) and circular laser polarizations ($\Omega_{\sigma,>}$, $\Omega_{\sigma,<}$)   vs $\theta$, for $E_L=1\times10^6$~V/m (linear) or $E_L=1\times10^6/\sqrt{2}$~V/m (circular). The circles show the case of Fig.~\ref{figure 4}(b).}
  \label{figure 4}
\end{figure}
In certain applications, one seeks optical coupling strengths that exceed the decay and dephasing rates of the involved initial and Rydberg states, such as, for instance, in blue shielding~\cite{secker, wang} and in pulsed excitation of RAIM wavepackets. When the initial state is strongly coupled to the Rydberg state, such that Rabi frequencies exceed decay and dephasing rates, we utilize a dressed-atomic-state picture~\cite{note1} to calculate PECs. In this case, a diagonalization of the full Hamiltonian in Fig.~\ref{figure 2}(b) is performed, because all levels are coherently coupled to each other. \par

In Fig.~\ref{figure 4}(a), we show the results for RAIM PECs of Rb as a function of internuclear distance $R$ for a case of strong optical coupling between the initial states $\{ \ket{5D_{3/2}, m_g} \}$ and the RAIM states. The PA laser is $\sigma^-$-polarized, has an intensity of $\approx 1.33\times10^9$~W/m$^2$ corresponding to a field amplitude of $E_L=1\times10^6$~V/m, and  propagates at an angle of $\theta = 45^{\circ}$ along the internuclear axis. This is a case in which all $\Omega$-couplings visualized in Fig.~\ref{figure 1}(c), between all pairs of $m_J$ and $m_g$, are allowed. The RAIM system is, in general, driven into a coherent superposition of states from different $m_J$ subspaces. The $m_J$-mixing is, however, effective only in the immediate vicinity of the crossings between dressed initial-state levels and RAIM PECs, such as within the small red box in Fig.~\ref{figure 4}(a). Outside these regions, the optical coupling is ineffective, and the PECs are not $m_J$-mixed. \par

In Fig.~\ref{figure 4}, we consider a case in which a $J_g = 3/2  \longleftrightarrow J = J_g-1 = 1/2$ optical coupling is the dominant drive, as seen from the energy location of the initial-state dressed levels relative to the Rydberg $45P_J$ states. In this case, the presence of dark (uncoupled) states is inevitable. In Fig.~\ref{figure 4}(b), we provide a zoom-in of the relevant intersection region marked by the red box in Fig.~\ref{figure 4}(a). The magnification clearly shows that there are four optically coupled levels and two dark states. The uncoupled states exist for any choice of $\theta$. 

The optical coupling strengths are evident from the splittings between the repelling dressed-state pairs in Fig.~\ref{figure 4}(b). The optical Rabi frequencies are given by the minimal values of the level splittings as a function of $R$, as indicated by the arrows in Fig.~\ref{figure 4}(b). The Rabi frequencies depend on the laser polarization and the angle $\theta$ between the molecular axis and the laser frame's $z$-axis. The latter is parallel to the laser electric field for linear and parallel to the laser propagation direction for circular polarizations. In Fig.~\ref{figure 4}(c) we show the Rabi frequencies for the same laser intensity as in Fig.~\ref{figure 4}(b), for the indicated laser polarizations versus  $\theta$. The Rabi frequencies for linear and circular polarizations exhibit opposite trends as a function of $\theta$, and there are no angles at which the optical coupling vanishes. For linear polarization, the pairs of optically-coupled dressed states have degenerate Rabi frequencies. It is further seen that for circular polarization the dark states are very weakly optically coupled, a behavior that is attributed to small admixtures of $\vert m_J \vert > 1/2$ states in the optically excited PECs.

\begin{figure}[t!]
 \centering
  \includegraphics[width=0.40\textwidth] {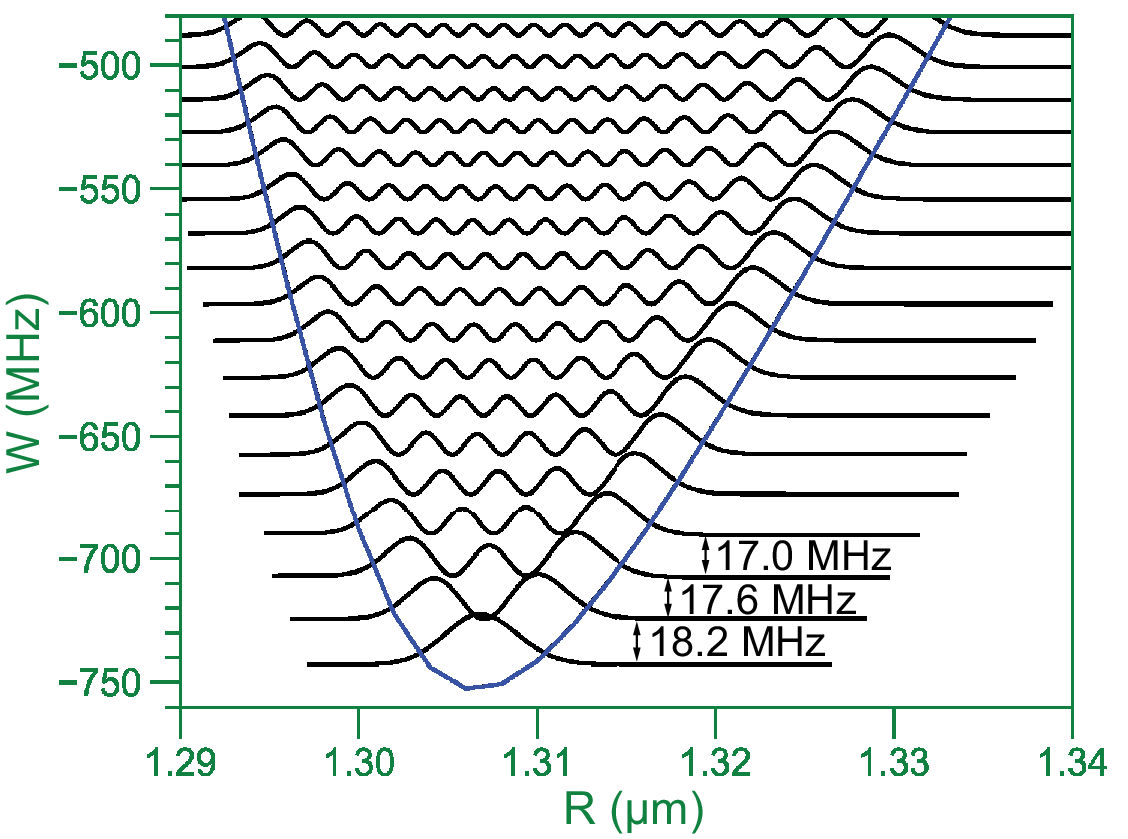}
    \caption{A magnified view of the lowest PEC minimum in the box in Fig.~\ref{figure 4}(a). The lowest 18 vibrational states in this PEC well are shown.}
  \label{figure 5}
\end{figure}

\subsection{Vibrational states}
\label{subsec:vibrstatesres}

The wells within PECs are typically several GHz deep, such as in Fig.~\ref{figure 4}(a), and support many vibrational states. This makes RAIMs attractive for laser-spectroscopic study and for time-dependent wave-packet studies. \par  

In Fig.~\ref{figure 5}, we show a magnified region of the lower PEC well associated with $45P_J$, outlined by the green box in Fig.~\ref{figure 4}(a), together with the squares of the vibrational wavefunctions for the lowest 18 vibrational states. The vibrational level structure is that of a perturbed harmonic oscillator in which the vibrational frequency intervals follow $f_n =  f_0 + \alpha_1 \nu + \alpha_2 \nu^2 ...$, with the lowest frequency interval $f_0$, the index $\nu =0, 1, 2...$, and only the lowest two dispersion terms shown. Fitting the lowest ten levels in Fig.~\ref{figure 5}, one finds $f_0=18.2$~MHz, $\alpha_1=-0.61$~MHz and $\alpha_2=0.020$~MHz for the first- and second-order dispersion coefficients. This finding could be tested, in the future, via laser spectroscopy. \par

This structure also lends itself to studies of vibrational wavepackets of RAIMs, including collapse and quantum revivals of wavepackets~\cite{eberly1980, robinett2004}. We use the results in Fig.~\ref{figure 5} to estimate the dephasing and revival times for vibrational wavepackets near the bottom of the potential well. Neglecting second- and higher-order dispersion, the dephasing time is $\sim 0.5 /(N \vert \alpha_1 \vert)$ and the revival time  $\approx 1 / \vert \alpha_1 \vert$, with the above lowest-order dispersion coefficient $\alpha_1$, and with the number of coherently excited states $N$. For wavepackets ranging in bandwidth from $N=5$ to 10 vibrational states, the dephasing time ranges from about about 300~ns to 700~ns, while the bandwidth-independent revival time is $\approx 1.6~\mu$s. These times are shorter than the RAIM lifetime, which should scale with the Rydberg-atom decay time (estimated at $\sim 100~\mu$s), and should therefore be observable. A detailed theoretical study on the dynamics of vibrational RAIM wavepackets can be an interesting topic for future investigations. These could also extend to the rotational structure of RAIM. We estimate the frequency interval of rotational energy levels to be in the range of 1-10~kHz, which would require quite sophisticated laser setups to be spectroscopically resolved, as well as lower ion velocities.

The stability of RAIM vibrational states as shown in Fig.~\ref{figure 5} may be limited by non-adiabatic coupling between the PECs. An initial estimation can be made based upon the Landau-Zener (LZ) formula~\cite{LZener}. The gaps at the avoided crossings between the RAIM PECs and neighboring PECs in Fig.~\ref{figure 3} indicate coupling strengths $a \sim h \times 150$~MHz and differential slopes of the PECs of $\Delta F \sim 10^{-17}$~N, while the vibration velocity for the lowest RAIM states in Fig.~\ref{figure 5} is $v \sim 0.3$~m/s. The LZ tunneling probability $P_d = \exp(- 2 \pi \Gamma)$ with $\Gamma = a^2/\vert \hbar \, v \, \Delta F \vert$ then is $< \exp(-100)$, which is favorable. We stress that the conditions for applicability of the LZ model are very poorly satisfied. Improved estimates of non-adiabatic decay rates require simulations of the vibrational quantum dynamics on multiple coupled PECs, which is beyond the scope of the present work.    

\subsection{Collisions on Rydberg-atom-ion PECs}
\label{subsec:coldcollisions}

The RAIM formalism for PECs in the strong optical coupling regime, described in the present paper, also directly applies to an analysis of certain optical shielding methods for atom-ion collisions. With the following examples we exhibit the close relationship between RAIMs and shielding of atom-ion collisions via strong optical coupling to Rydberg states.

\begin{figure}[b!]
 \centering
  \includegraphics[width=0.49\textwidth]{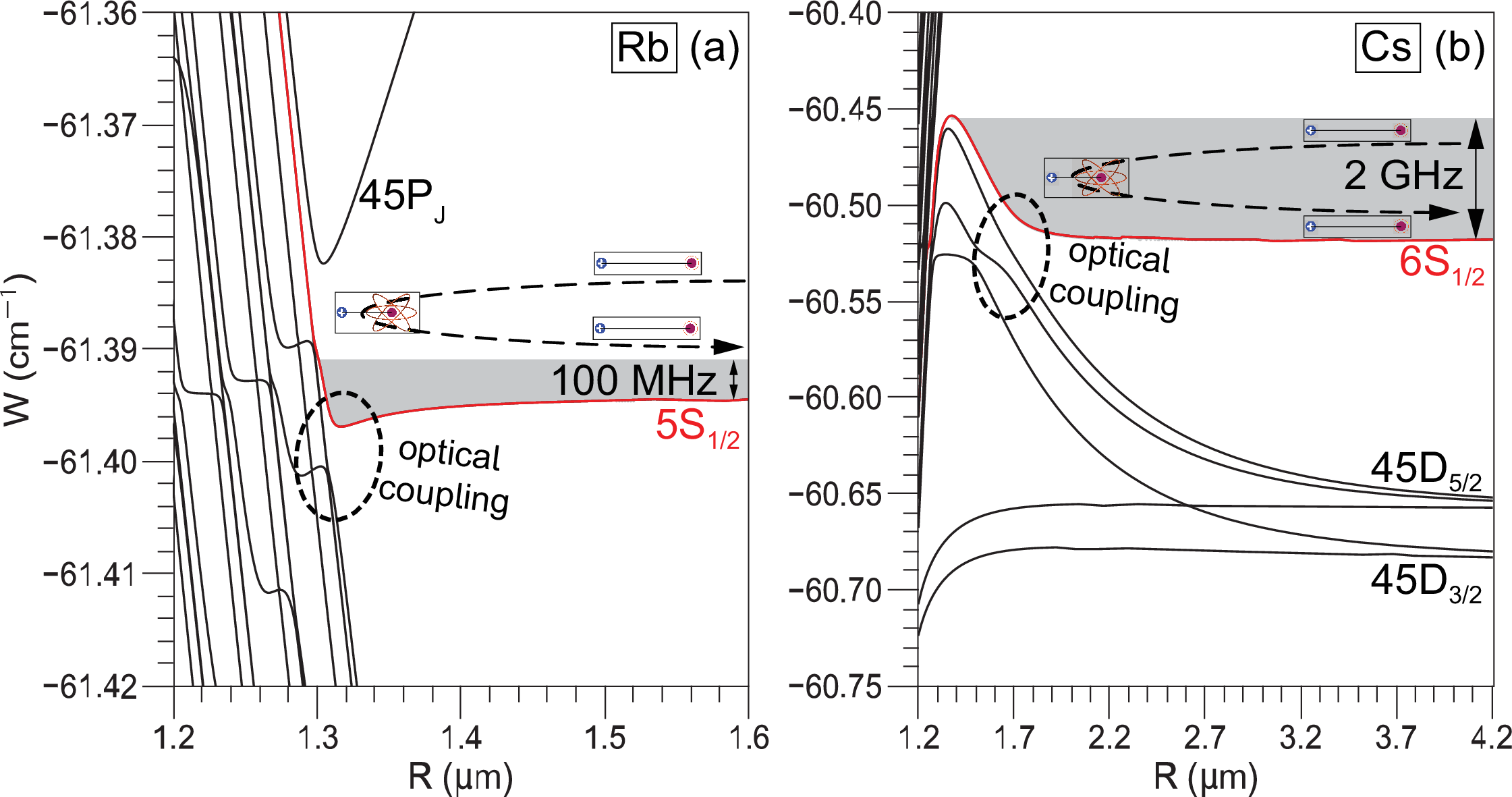}
  \caption{(a) Soft potential barrier for Rb atom-ion collisions, affected by strong optical coupling of the atomic ground state to Rydberg PECs slightly below the 45$P_J$ states. The barrier is $\sim h \times 100$~MHz high. (b) Soft potential barrier for Cs atom-ion collisions, affected by strong optical coupling of the atomic ground state to blue-shifted PECs with 45$D_{5/2}$-character. The barrier is $\sim h \times 2$~GHz high.}
  \label{figure 6}
\end{figure}

As one example, lowering the dressed-initial-state detuning in Fig.~\ref{figure 4}(a) from -61.36~cm$^{-1}$ to -61.393~cm$^{-1}$, using an initial state of $5S_{1/2}$, a laser intensity $\approx2.12\times10^{12}$~W/m$^2$ (electric-field amplitude of $E_L=4 \times 10^7$~V/m), $\theta=45^\circ$, and linear polarization, results in the PEC landscape shown in Fig.~\ref{figure 6}~(a). In a cold-collision scenario, the configuration introduces a soft energy barrier at about 1.3~$\mu$m that is about $h \times$100~MHz high. In cold-atom systems, the barrier can be used to prevent deep, inelastic collisions between Rb atoms and ions, as described in detail in~\cite{secker, wang} for a lower Rb $nP_J$ Rydberg level.\par

In Cs, efficient shielding of inelastic atom-ion collisions can likely be achieved using the blue-shifting $m_J = 1/2$ Stark component of $nD_{5/2}$, which is fairly unique among alkali Rydberg atoms due the fact that the $P$- and $D$ quantum defects both have similar non-integer parts near 0.5. The proximity of the $nP$ and $nD$ levels to each other results in strong level repulsion in the ion electric field, with $nD$ levels shifting up (see Fig.~\ref{figure 6}(b)). At the same time, the field-free $nD$ levels are quite distant from the next-higher hydrogenic manifold, allowing large blue shifts of the $nD$ levels before mixing with hydrogenic levels occurs. Interplay of these effects results in unusually high potential barriers in Cs, conducive to efficient collisional shielding. As an example, in Fig.~\ref{figure 6}~(b) we employ the RAIM-PEC formalism to find the PECs for collisions between Cs atoms and ions with the coupling laser tuned above the atomic $45D_{5/2}$ resonance. The initial (dressed) atomic state $6S_{1/2}$ is tuned to -60.52~cm$^{-1}$, the laser field is linearly polarized and has an intensity $\approx3.32\times10^{12}$~W/m$^2$ (electric-field amplitude of $E_L=5 \times 10^7$~V/m), and $\theta=45^\circ$. The configuration introduces a comparatively tall, soft collisional barrier that is located at about 1.7~$\mu$m and is about $h \times$ 2~GHz high. 

\section{Experimental considerations}
\label{sec:expt}

In the following we discuss two possible approaches to study RAIMs. The first, and arguably easier, method is to perform high-precision spectroscopy measurements with quasi-continuous narrow-band ($\leq 1$ MHz) lasers. The second method is to excite vibrational wavepackets with pulsed lasers in the strong optical coupling regime discussed in Section~\ref{subsec:vibrstatesres}. The first approach requires simpler laser setups (several diode lasers should suffice), and it could be used for proof-of-principle experiments for RAIM formation. The second approach has great experimental potential, as it can be used to explore the vibrational dynamics of the molecules.

\subsection{Laser spectroscopy with narrow-band lasers}
\label{subsec:dstateprep}

\begin{figure}[t!]
 \centering
  \includegraphics[width=0.44\textwidth]{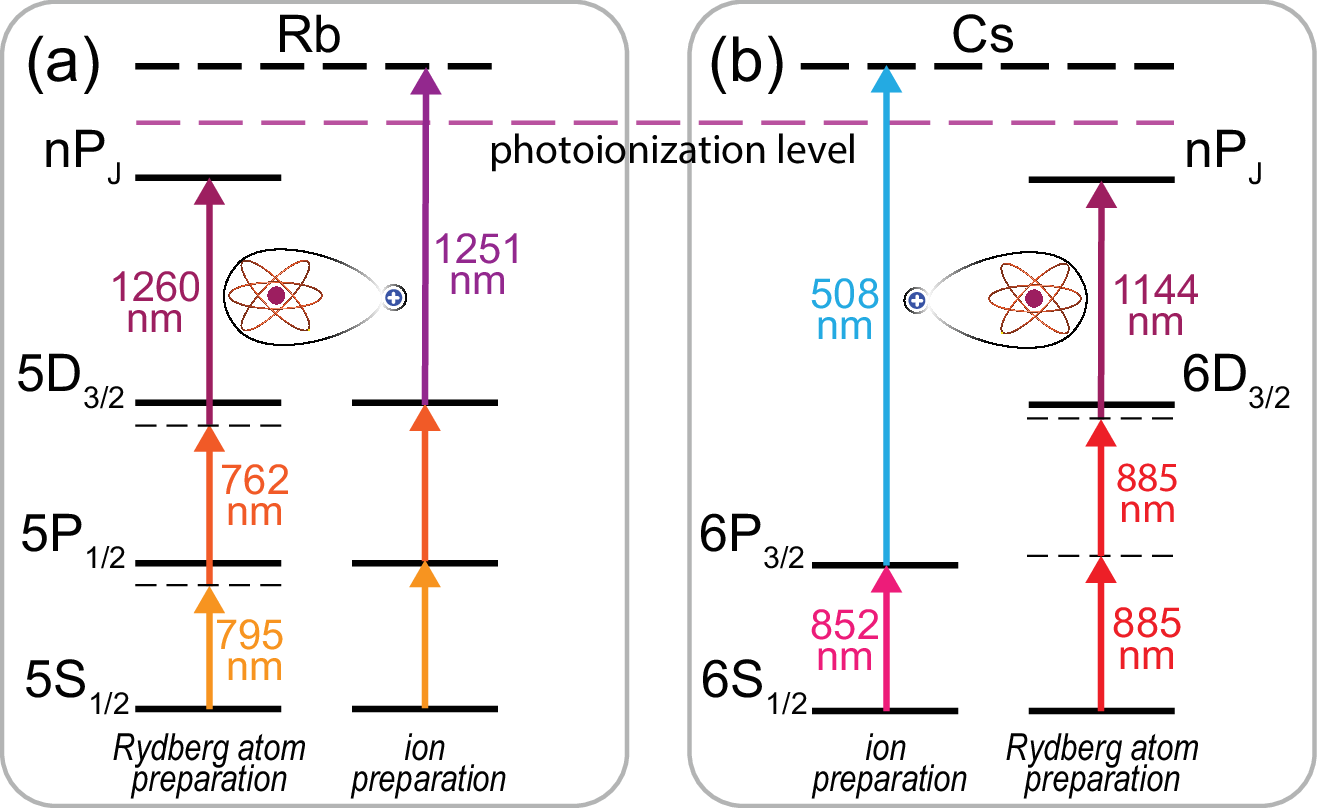}
  \caption{(Color online) Energy level diagram (not to scale) with relevant wavelengths for the experimental realization of RAIM with Rb (a) and Cs (b). The lower stages of the Rydberg-atom transitions are driven off-resonantly. See text for details.}
  \label{figure 7}
\end{figure}

In this subsection, we discuss experimental methods to prepare RAIMs of Rb or Cs via high-resolution laser spectroscopy with quasi-continuous lasers. For Rb, some details are visualized in Fig.~\ref{figure 7}(a). The $5S_{1/2} \rightarrow nP_J$ ground-to-Rydberg direct transition is less frequently used than multi-photon schemes, as it would require a laser with a wavelength of $\sim 300$~nm. A method of preparing Rb atoms in Rydberg $P$-states with an all-infra-red laser system is to first excite atoms via a low-lying $D$-state into Rydberg $P$-states. For instance, in Rb the following three-step excitation can be used: $5S_{1/2} \rightarrow 5P_{1/2}$ at 795~nm, $5P_{1/2} \rightarrow 5D_{3/2}$ at 762~nm, $5D_{3/2} \rightarrow nP_{J}$ at 1260~nm. These three wavelengths can be attained with conventional diode lasers and without frequency-doubling, and the Rydberg-excitation stage has relatively large excitation matrix elements and rates (see Fig.~\ref{figure 2}). The atoms are photo-associated into RAIMs by seeding the atom cloud with ions that have a kinetic energy low enough to avoid line broadening by more than a fraction of the vibrational energy level spacing. Due to ion recoil, the photo-ionization (PI) energy in Rb should be within about 100~GHz to 1~THz above the PI threshold in order to be able to resolve vibrational states with spacings as in Fig.~\ref{figure 5}. This corresponds to wavelengths of 0.5~nm to 5~nm less than the Rb$~5D_{3/2}$ PI wavelength of 1251.52~nm. \par

A corresponding level diagram and excitation scheme for Cs is displayed in Fig.~\ref{figure 7}(b). Two-photon excitation at 885~nm, utilized in previous experiments~\cite{Ohtsuka2005, ChenTJ2018}, can be used to prepare Cs atoms in the $6D_{3/2}$ state. Then, a laser at 1144~nm can be used to photo-associate $nP_J$ RAIMs of Cs. At the same time, the excitation region can be seeded with low-energy ions by PI of atoms in the 6$P_{3/2}$ state, accessed via the 852-nm Cs D2 line, and a 508-nm PI laser. To keep the ion kinetic energy low enough to be able to resolve vibrational levels, the upper-laser wavelength should be between 0.2 and 2~nm below the  $6P_{3/2}$ PI threshold, which is at 508.28~nm. \par

The ions must generally be prepared at a low enough density to avoid Coulomb acceleration and expansion before and during the photo-association pulse. Other options that could be explored to reduce ion energy include using an ion trap to prepare laser-cooled ions. Ion velocities in the range $\lesssim 0.2$~m/s, as proposed here, translate into photo-association times between 100~ns and 1~$\mu$s, leading to interaction-time broadening of less than 10~MHz, which is sufficiently low to resolve vibrational structure.\par 

Figure~\ref{figure 2}(b) suggests RAIM production rates on the order of 10$^5$~s$^{-1}$, which, for PA time windows in the range of hundreds of nanoseconds, suggests a RAIM production probability in the range of 3$\%$. Using an experimental cycle rate of 100~Hz~\cite{jamie}, assuming several tens of ions embedded in a dense atomic vapor, and estimating an ion detection efficiency of 30$\%$, the count rate would be several 10~s$^{-1}$. Considering that the signal would be virtually free of background counts from distant atomic lines, laser-spectroscopic RAIM measurements appear quite feasible. \par

It is worth noting that mixed-species RAIMs between a Rb or a Cs Rydberg atom and any point-like positive ion have structures similar to the ones considered in this paper, with different vibrational spacings due to the different effective masses. This includes commonly used clock ions such as Hg$^+$, Sr$^+$, Yb$^+$, etc. Since laser-assisted cooling of negative ions has been demonstrated recently~\cite{cerchiari}, we have also studied RAIMs between negative ions and Rydberg atoms. Qualitatively, these do not differ greatly from RAIMs with positive ions. However, quantitatively they have noticeably different PECs, because the Rydberg-electron charge distribution of the adiabatic Rydberg states on their PECs are shifted away from the ion, into regions of diminishing electric field. This contrasts to the case of positive ions, where the Rydberg-electron charge distribution is pulled towards the ion, into regions of increasing electric field. The resultant differences in PECs and RAIM states could be a subject of a future study. \par

\subsection{Vibrational wavepackets}
\label{subsec:wavepacketexp}

In Section~\ref{subsec:vibrstatesres}, we have provided estimates for dephasing and recurrence times of vibrational wavepackets consisting of 5-10 vibrational levels of $45P_J$ RAIMs. This number of vibrational levels is large enough to form reasonably well-localized wavepackets, and it is small enough to avoid the effect of higher-order dispersion terms in the vibrational energy-level series. As seen in Fig.~\ref{figure 5}, such a wavepacket would cover a spectral bandwdith of 100-200~MHz. Its realization requires a transform-limited laser pulse with a duration of about 5-10~ns, corresponding to the aforementioned spectral bandwidth. A shorter (longer) laser pulse leads to the excitation of more (fewer) vibrational levels and shorter (longer) wavepacket dephasing times. Considering only lowest-order dispersion, the revival time is independent of the number of vibrational levels, and the revival remains strong even for large numbers of coherently excited vibrational states. \par

We concentrate on Rb RAIMs to further discuss the relevant requirements for the laser. The excitation scheme for Rb RAIMs suggested in Sec.~\ref{subsec:dstateprep} would require either continuous-wave or pulsed lower-transition lasers and a high-intensity 1260-nm laser pulse. Presently, there are no convenient choices for 1260~nm transform-limited pulsed lasers with pulse durations in the range of 5-10~ns. An alternative approach would be to excite the atoms directly from the $5S_{1/2}$ ground state into $nP$ Rydberg state. In this case, the excitation pulse would be centered at a wavelength of 297~nm, which could be achieved by frequency-doubling of a pulsed Rhodamine~6G dye laser pumped with a 532-nm Nd:YAG laser. The frequency-doubling could be achieved, for instance, with a beta barium borate or a lithium niobate crystal. \par

The wavepacket dynamics can be probed using pump-probe schemes, the most basic version of which would be to apply two coherent PA pulses with a variable delay time and to measure the RAIM yield versus delay time~\cite{Fatemi2001, McCabe2009}. Additional experiments, e.g., using quantum gas microscopy~\cite{hollerith, hollerith2020}, could be employed for spatio-temporal mapping of the RAIM vibrational wavefunctions. Alternatively, the wavefunctions could be studied in correlation measurements. In this scheme, the RAIMs would be field-ionized and imaged with a single-ion-resolving imaging system. Similar methods have been used to study correlations in both Rydberg atom and ion plasma systems~\cite{Thaicharoen2016, Viray2020}.

\section{Conclusion}
\label{sec:conclusion}

We have shown through PEC calculations that it is possible to form Rydberg atom-ion molecules that are bound by way of multipolar interactions between the two particles. We have provided detailed results on several cases in which a Rb or Cs neutral-atom initial state is optically coupled to Rydberg-state manifolds. We have also discussed potential experimental realizations of RAIMs, differences in forming the molecules with rubidium versus with cesium, and the possibility of preparing vibrational wavepackets in these molecules. Modern state-of-the-art experimental setups with ultracold Rydberg atoms and trapped ions already possess all necessary components to form and observe these molecules. It may be anticipated that the observation of RAIMs will open a new direction in the field of ultralong-range Rydberg molecules and many-body physics.

\section*{ACKNOWLEDGMENTS}
This work was supported by NSF grant Nos. PHY-1506093 and PHY-1806809. X.H. acknowledges support from the China Scholarships Council (No. 201808140193). J.Z. acknowledges support from the National Natural Science Foundation of China (NSFC) (61835007).

\bibliography{references.bib}

\begin{thebibliography}{78}
\bibitem [{not()}]{note1}%
  \BibitemOpen
  \href@noop {} {}\bibinfo {note} {We are using the term ``dressed atomic
  state'' to identify an interaction-picture atomic state in a classical laser
  field.}\BibitemShut {Stop}%
\end{thebibliography}

\end{document}